\documentclass[pra,twocolumn,showpacs,groupedaddress]{revtex4}

\usepackage{amsmath}
\usepackage{amsfonts}
\usepackage{stmaryrd}
\usepackage{amssymb}
\usepackage{graphics,epsfig}

\begin{document}

\title{Quantum interference in a four-level system of a $^{87}\mathrm{Rb}$ atom:
\\ Effects of spontaneously generated coherence}

\author{Dongsheng Wang}
\email{wdsn1987@gmail.com}\affiliation{School of Physics, Shandong
University, Jinan 250100, China}

\author{Yujun Zheng}
\email{yzheng@sdu.edu.cn} \affiliation{School of Physics, Shandong
University, Jinan 250100, China}

\date{November 3, 2010}

\begin{abstract}
In this work, the effects of quantum interference and spontaneously
generated coherence (SGC) are theoretically analyzed in a four level
system of a $^{87}\mathrm{Rb}$ atom. For the effects of SGC, we find
that a new kind of EIT channel can be induced due to destructive
interference, and the nonlinear Kerr absorption can be coherently
narrowed or eliminated under different strengths of the coupling and
switching fields.
\end{abstract}


%42.50.Gy Effects of atomic coherence on propagation, absorption,
%    and amplification of light; electromagnetically induced transparency and absorption
%42.65.Hw Phase conjugation; photorefractive and Kerr effects
%42.50.Ar Photon statistics and coherence theory

\maketitle

\section{INTRODUCTION}
\label{sec:INTRODUCTION}

The quantum interference and coherence play the central role in the
fields of nonlinear optics, quantum information processing, and
quantum wells/dots etc. In recent years, the processes of
electromagnetically induced transparency (EIT)
\cite{boller,harris,flei,merriam,chong,lezama,yanpeng,buth,aaa} are
widely studied in various systems. Many processes based on EIT, for
example, the electromagnetically induced absorption (EIA)
\cite{lezama}, the EIT based four-wave and six-wave mixing
\cite{yanpeng}, the EIT for X-rays \cite{buth}, and the EIT on a
single artificial atom \cite{aaa} are studied. Also, the giant Kerr
effect (GKE) \cite{schmidt,myin,minyan,kang,braje,jiang,shujing,
bajcsy,sinclair1,sinclair2,gong,imamoglu,gao0,xihua,minxie,plenio,milburn}
is the natural application of EIT by introducing the switching
(signal) field, which has many applications, e.g., in the photonic
Mott insulator \cite{plenio} and the circuit quantum electrodynamics
\cite{milburn} etc. Physically, once the dark state within the EIT
is destroyed by the disturbance of the switching field, the EIT
disappears, the nonlinear susceptibility and absorptive photon
switching occur. The processes of EIT and GKE could be changeable by
turning the switching field on and off, and the photon process
within can be managed. Recently, the effects of spontaneously
generated coherence (SGC) (or termed as vacuum-induced coherence
(VIC)) stimulate one's interests. For instance, in three-level
systems the nonlinear Kerr absorptions are shown to be greatly
enhanced via SGC \cite{gong}. For four-level system, the effects of
SGC are studied in the $N$-type system \cite{jiang} on the nonlinear
Kerr absorption, and in double $\Lambda$-type system on the coherent
population transfer \cite{xihua,minxie}.

Experimentally, it is effective to simulate the effects of SGC by
changing the property of the vacuum \cite{patnaik,patnaik2}, and/or
extra driving fields, such as a dc field
\cite{ficek,berman,agarwal2}, a microwave field \cite{jiahua}, or a
laser field \cite{cwang,gao} etc. Some experimental groups have
realized part effects of SGC in the $V$-type, four-level
$\Lambda$-type, and other systems recently
\cite{gao,cwang,xia,xwang,norris}.

In the present work we study the effects of SGC in EIT and GKE in
the four-level of $N$-type and double $\Lambda$-type systems based
on the generating function approach developed recently
\cite{zheng,zhengj,mukamel,hey,bel06,peng,peng09,penga,han1,han2,pengwang,dwang}.
In the double $\Lambda$-type system, SGC can cause new EIT channels
due to destructive interference, which could be termed as
``vacuum-assisted transparency'' (VAT). The usual EIT requires that
the detunings of the coupling and probe fields satisfy the
two-photon resonance condition. With the effects of SGC, however,
transparency can be obtained beyond the two-photon process. This
means that we can get the transparency window within the whole
detuning range of the probe field. For the nonlinear Kerr absorption
in the $N$-type system, the spectra can be narrowed or eliminated
with the effect of SGC, under different strengths of the coupling
field $\Omega_c$ and switching field $\Omega_s$. We demonstrate that
the destroyed dark state can be repaired via SGC under some
conditions. Also, we investigate the photon switching process via
photon statistics, the variance of the photon distribution can be
presented by Mandel's $Q$ parameter. One of the important case is $Q
< 0$, which shows anti-bunching when photon switching occurs.

For the four-level systems in this work, we demonstrate that SGC
plays different roles in the linear and nonlinear absorption. By
introducing SGC, the ability of the original EIT and GKE can be
improved, also the photon statistics can be greatly affected and
reflect the characters of different dynamics. The reasons we
investigate two type systems together are as follows. Firstly, the
two systems are, actually, the same with each other. They have the
same structures, and a $^{87}\mathrm{Rb}$ atom can be their concrete
system. Secondly, the $N$-type and double $\Lambda$-type systems are
apparent type, their difference comes from the numbers of driving
laser fields.

SGC (VIC) shows us another pathway of quantum interference, which
serves to mediate the relative contribution of the dipole
transitions, and represents the directly coupling between the
coherence of the system. SGC (VIC), represented by the generalized
decay constants (GDCs), can lead to some novel features in dynamics
of atomic systems, and it is shown that SGC plays the important role
in many phenomenons \cite{agarwal3,car1,car2,fleischhauer,
javanainen,Shore,zhusy2,agarwal4,zhu,knight,zhoupeng1,zhoupeng2,anton1,anton2,anton3,
patnaik,patnaik2,berman,agarwal2,ficek,jiahua,gao,cwang,xia,xwang},
such as the dark state resonance \cite{fleischhauer,car1,car2},
coherent population trapping and transfer
\cite{javanainen,zhusy2,agarwal4,Shore}, spectral narrowing and
elimination \cite{zhu,zhoupeng1,knight}, gain without population
inversion \cite{zhoupeng2}, spectrum squeezing
\cite{anton1,anton2,anton3} etc.

This paper is organized as follows. In Section~\ref{sec:TF}, we give
our theoretical framework of the quantum dynamics of four-level
system of a $^{87}\mathrm{Rb}$ atom. In our theoretical framework,
we include the generalized decay constants (GDCs) within the
rotating wave approximation (RWA). In Section~\ref{sec:NR}, we
present our theoretical results of the effect of SGC. In
Section~\ref{sec:CL}, we give brief conclusion and discussion.

\begin{figure}%[b]
\includegraphics[scale=0.25]{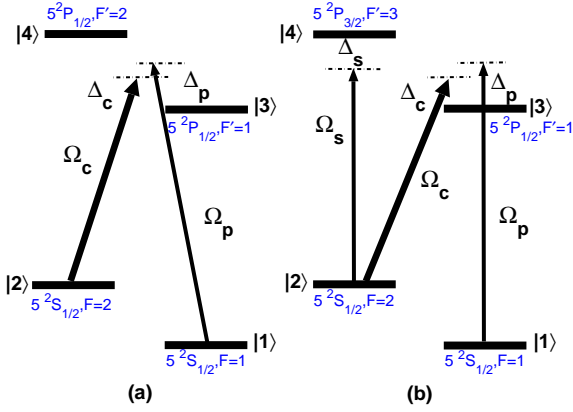}
\caption{(Color online) The schematic diagram of four-level system
in this work. Panel (a) is double $\Lambda$-type, and panel (b) is
$N$-type. The proposed $D$-line energy levels of the
$^{87}\mathrm{Rb}$ atom are marked on the diagram.}
\label{fig.schematic}
\end{figure}

\section{Theoretical FRAMEWORK}
\label{sec:TF}

In this work we theoretically study the quantum dynamics of a
four-level quantum system driven by external laser fields. The
four-level systems are shown in Fig.~\ref{fig.schematic}. In our
study, we consider the transition dipole moments: $| 3 \rangle \to |
1 \rangle$, $| 3 \rangle \to | 2 \rangle$, $| 4 \rangle \to | 1
\rangle$, and $| 4 \rangle \to | 2 \rangle$. However the direct
transitions between the states $| 2 \rangle$ and $| 1 \rangle$, and
$| 4 \rangle$ and $| 3 \rangle$ are dipole forbidden. We consider
the evolution of the reduced density matrix of the system $\sigma(t)
\equiv \textrm{Tr}_{R}\left\{\rho(t)\right\}$. The evolution of the
reduced density matrix satisfies the Liouville equation
\cite{blum,ct,lmandel,milonni}
\begin{equation}
\dot{\sigma}_{ij}(t)=\mathcal{L}_{ij, kl}\sigma_{kl},
\end{equation}
where $\mathcal{L}_{ij, kl}$ is the Liouville super-operator.
We study the effects of SGC on linear and nonlinear absorptions and also on the photon statistics.
One of the convenient approach to calculate photon counting moments is generating function approach
developed recently \cite{zheng,zhengj,mukamel,hey,bel06,peng09,penga,han1,han2,pengwang,peng,dwang}.
The generating function is defined as \cite{zheng,zhengj}
\begin{equation}
\mathcal {G}(s,t)=\sum_{n=0}^{\infty}\sigma^{(n)}(t)s^{n},
\end{equation}
where $s$ is called the counting variable, $\sigma^{(n)}(t)$ describes the probability
that $n$ photons have been emitted in the time interval $[0,t]$, and
\begin{equation}
\sigma(t)=\sum_{n=0}^{\infty}\sigma^{(n)}(t).
\end{equation}

Similar with Refs.~\cite{zheng,zhengj,mukamel,hey,bel06,peng09,penga,han1,han2,pengwang,peng,dwang},
the evolution of the generating function of four-level system can,
under the rotating wave approximation (RWA), be formally written as
\begin{eqnarray}
\label{eq:gf} \dot{\mathcal{G}}_{ij}(s,t) &=&-i\omega_{ij}\mathcal{G}_{ij}-
              i/2\sum_{m}(\Omega_{mj}\mathcal {G}_{im}-\Omega_{im}\mathcal{G}_{mj}) \\ \nonumber
   &  &  +2s\sum_{k>i,l>j}\gamma_{kijl}\mathcal{G}_{kl}-
         \sum_{k>m}\gamma_{immk}\mathcal{G}_{kj}-\sum_{l>n}\gamma_{lnnj}\mathcal {G}_{il} ,
\end{eqnarray}
where the energy spacing $\omega_{ij} = \omega_i-\omega_j$, the Rabi
frequency $\Omega_{ij}(t) =
\bm{\mu}_{ij}\cdot\bm{\mathcal{E}}(t)/\hbar $, $\bm{\mathcal{E}}(t)$
is the polarization of the field with frequency $\omega_L$, the GDCs
are
$\gamma_{ijkl}=\bm{\mu}_{ij}\cdot\bm{\mu}_{kl}\frac{|\omega_{kl}|^3}{6\varepsilon_0\hslash
\pi c^3}$, where $\omega_{kl} \approx \omega_{ij}$ has been taken
into account. $k>i$ represents that the level $|k\rangle$ sits above
the level $|i\rangle$. Eq.~(\ref{eq:gf}) for the double
$\Lambda$-type system and $N$-type system considered in this work is
explicitly presented in the following section.

In our study, the GDCs $\gamma_{ijkl}$ can be expressed as \cite{xihua,anton1,anton2,anton3,zhoupeng2,ficek}
\begin{equation}
\gamma_{kiil}=\beta \sqrt{\gamma_{kiik}\gamma_{liil}},
\end{equation}
the parameter $\beta$ could be named as the ``SGC factor'', it takes
$0\leq \beta \leq 1$. When the dipole moments are perpendicular
$\beta=0$, and $\beta=1$ for the dipole moments parallel case.

In order to extract the information of photon emission events, we
define the working generating function

\begin{equation}
\mathcal {Y}(s,t)=\sum_{k=1}^{4}\mathcal {G}_{kk}(s,t).
\end{equation}

The factorial moments $\langle N^{(m)}\rangle(t)$ and emission
probability of $n$ photon $P_n(t)$ can be obtained via \cite{zheng,zhengj}
\begin{eqnarray}
\langle N^{(m)}\rangle(t) &=& \langle N(N-1)(N-2) \cdots
(N-m+1)\rangle(t)\\\nonumber
       &=& \frac{\partial^{m}}{\partial s^{m}}\mathcal{Y}(s, t)|_{s=1} ,
\end{eqnarray}
and
\begin{equation}
P_n(t) = \frac{1}{n!}\frac{\partial^{n}}{\partial
s^{n}}\mathcal{Y}(s,t)|_{s=0} .
\end{equation}

Correspondingly, the absorption line shapes and  Mandel's $Q$
parameter can be obtained
\begin{equation}
I (\omega_L) = \frac{d}{dt}{\langle N^{(1)}
\rangle}(t)|_{t\rightarrow \infty} ,
\end{equation}
\begin{equation}
Q (\omega_L)= \frac{\langle N^{(2)}\rangle-\langle N^{(1)}
\rangle^{2}}{\langle N^{(1)} \rangle}.
\end{equation}

\section{Numerical Results}
\label{sec:NR}

In this section, we study the effects of SGC on electromagnetically
induced transparency and nonlinear Kerr absorption, and on the
photon statistics. In our numerical results, we plot absorption line
shapes and Mandel's $Q$ parameter. The absorption is monitored via
fluorescence of spontaneously emitted photons. The reason is that
the correspondence between photon emission and absorption is insured
by conservation of energy \cite{peng09,bel06}. Our results can be
measured experimentally based on a $^{87}\mathrm{Rb}$ atom. For the
convenient of future experiment, all the parameters used in our
numerical results are the parameters of the $^{87}\mathrm{Rb}$ atom,
the proposed $D$-line energy levels of the $^{87}\mathrm{Rb}$ atom
in this work are also marked in Fig.~\ref{fig.schematic}. In the following equations of generating
function (\ref{eq:2lambda}) and (\ref{eq:n4}), for convenience, we
use the shorthand
\begin{eqnarray}
\gamma_{31}  &\equiv& \gamma_{3113}, \;\;\; \gamma_{32}  \equiv
\gamma_{3223}, \\\nonumber \gamma_{41}  &\equiv& \gamma_{4114},
\;\;\; \gamma_{42}  \equiv \gamma_{4224}, \\\nonumber \gamma_{314}
&\equiv& \gamma_{3114}, \;\;\; \gamma_{324} \equiv \gamma_{3224},
\\\nonumber \Gamma_{3}   &\equiv& \gamma_{3113}+\gamma_{3223},
\\\nonumber \Gamma_{4}   &\equiv& \gamma_{4114}+\gamma_{4224},
\\\nonumber \Gamma_{34}  &\equiv& \gamma_{3114}+\gamma_{3224}.
\end{eqnarray}
It should be noted that we normalize the absorption line shapes in
the following sections.

\subsection{DOUBLE $\Lambda$-TYPE SYSTEM}
\label{sec:2-2 SYSTEM}

\begin{figure}
\centering
\includegraphics[scale=0.2]{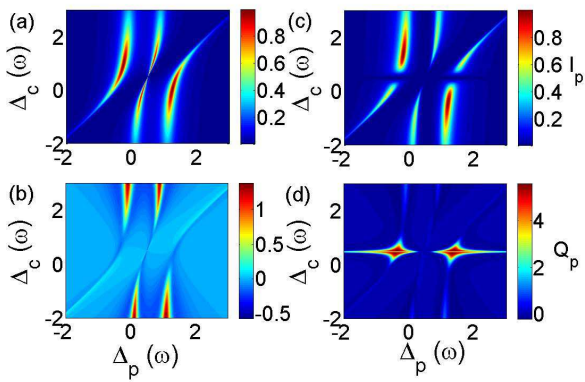}
\caption{(Color online) The absorption line shapes $I_p$ (panels (a)
and (c)) and Mandel's $Q_p$ parameter (panels (b) and (d)) of the
EIT of the double $\Lambda$-type system in the detuning space
$\Delta_p$-$\Delta_c$. The parameters are used the same with the
$^{87}\mathrm{Rb}$ atom \cite{steck,myin,sinclair2,maric,aum}:
$\gamma_{31}=\gamma_{32}=\gamma_{41}=\gamma_{42}=1.4375$MHz,
$\omega=814.5$MHz. The initial state is $|\psi\rangle$=$|1\rangle$.
The other parameters are $\Omega_p=0.1\omega$, $\Omega_c=\omega$,
and $\gamma_{314}=\gamma_{324}=0$ for $\beta=0$ (panels (a) and
(b));
$\gamma_{314}=\gamma_{324}=\sqrt{\gamma_{31}\gamma_{41}}=1.4375$MHz
for $\beta=1$ (panels (c) and (d)). In the study of double
$\Lambda$-type system, the detunings ($\Delta_p$, $\Delta_c$) and
Rabi frequencies ($\Omega_p$, $\Omega_c$) are in the units of
$\omega$.} \label{fig:EIT1}
\end{figure}

In this subsection, we study the process of EIT in the double $\Lambda$-type system.
Its schematic diagram is shown in panel (a) of Fig.~\ref{fig.schematic}.
In our numerical results, we refer to the parameters of the $^{87}\mathrm{Rb}$ atom,
namely, we set $|5^2S_{1/2}, F=1 \rangle=|1 \rangle$, $|5^2S_{1/2}, F=2 \rangle=|2\rangle$,
$|5^2P_{1/2}, F'=1 \rangle=|3 \rangle$, and $|5^2P_{1/2}, F'=2 \rangle=|4 \rangle$.
The parameters are listed in the caption of Fig.~\ref{fig:EIT1}.

In this system, the transitions of $| 3 \rangle \rightarrow | 1
\rangle$ and $| 4 \rangle \rightarrow | 1 \rangle$ are probed by the
weak probe field, the transitions of $| 3 \rangle \rightarrow | 2
\rangle$ and $| 4 \rangle \rightarrow | 2 \rangle$ are coupled by
the strong coupling field. The Rabi frequencies are defined as
$\Omega_{13}=\Omega_{14}=\Omega_p$,
$\Omega_{23}=\Omega_{24}=\Omega_c$. The detunings are defined as
$\Delta_p=\omega_p-\omega_{31}$, $\Delta_c=\omega_c-\omega_{32}$,
and the separation of the excited states $\omega=\omega_4-\omega_3$.
The generating function of Eq.~(\ref{eq:gf}) for the double
$\Lambda$-type system is obviously expressed as
\begin{widetext}
\begin{eqnarray}
\label{eq:2lambda}
\dot{\mathcal{G}}_{11}&=&2s(\gamma_{31}\mathcal{G}_{33}+\gamma_{314}\mathcal{G}_{34}+
     \gamma_{314}\mathcal{G}_{43}+\gamma_{41}\mathcal{G}_{44})
     -\frac{i}{2}\Omega_{p}(\mathcal {G}_{13}+\mathcal{G}_{14}-
     \mathcal{G}_{31}-\mathcal {G}_{41}),\\ \nonumber
\dot{\mathcal{G}}_{22}&=&2(\gamma_{32}\mathcal{G}_{33}+\gamma_{324}\mathcal{G}_{34}+
     \gamma_{324}\mathcal{G}_{43}+\gamma_{42}\mathcal{G}_{44})
     -\frac{i}{2}\Omega_{c}(\mathcal {G}_{23}+\mathcal{G}_{24}-
     \mathcal{G}_{32}-\mathcal {G}_{42}),\\ \nonumber
\dot{\mathcal{G}}_{33}&=&-2\Gamma_3\mathcal{G}_{33}-\Gamma_{34}(\mathcal{G}_{34}+
     \mathcal{G}_{43})-\frac{i}{2}\Omega_{p}(\mathcal{G}_{31}-\mathcal{G}_{13})-
     \frac{i}{2}\Omega_{c}(\mathcal {G}_{32}-\mathcal{G}_{23}),\\ \nonumber
\dot{\mathcal{G}}_{44}&=&-2\Gamma_4\mathcal{G}_{44}-\Gamma_{34}(\mathcal{G}_{34}+
     \mathcal{G}_{43})-\frac{i}{2}\Omega_{p}(\mathcal{G}_{41}-\mathcal{G}_{14})-
     \frac{i}{2}\Omega_{c}(\mathcal{G}_{42}-\mathcal{G}_{24}),\\ \nonumber
\dot{\mathcal{G}}_{12}&=&-i(\Delta_p-\Delta_c)\mathcal{G}_{12}-
      \frac{i}{2}\Omega_{c}(\mathcal{G}_{13}+\mathcal{G}_{14})+
      \frac{i}{2}\Omega_{p}(\mathcal{G}_{32}+\mathcal{G}_{42}), \\ \nonumber
\dot{\mathcal{G}}_{13}&=&-i\Delta_p\mathcal{G}_{13}-\Gamma_3\mathcal{G}_{13}-
      \Gamma_{34}\mathcal{G}_{14}-\frac{i}{2}\Omega_{p}(\mathcal{G}_{11}-\mathcal{G}_{33}-
      \mathcal {G}_{43})-\frac{i}{2}\Omega_{c}\mathcal {G}_{12},\\ \nonumber
\dot{\mathcal{G}}_{14}&=&-i(\Delta_p-\omega)\mathcal{G}_{14}-
      \Gamma_{34}\mathcal{G}_{13}-\Gamma_4\mathcal{G}_{14}-
      \frac{i}{2}\Omega_{p}(\mathcal{G}_{11}-\mathcal{G}_{34}-\mathcal {G}_{44})-
      \frac{i}{2}\Omega_{c}\mathcal {G}_{12},\\ \nonumber
\dot{\mathcal{G}}_{23}&=&-i\Delta_c\mathcal{G}_{23}-\Gamma_3\mathcal{G}_{23}-
     \Gamma_{34}\mathcal{G}_{24}-\frac{i}{2}\Omega_{c}(\mathcal {G}_{22}-
     \mathcal{G}_{33}-\mathcal{G}_{43})-\frac{i}{2}\Omega_{p}\mathcal{G}_{21},\\ \nonumber
\dot{\mathcal{G}}_{24}&=&-i(\Delta_c-\omega)\mathcal{G}_{24}-\Gamma_{34}\mathcal{G}_{23}-
     \Gamma_4\mathcal{G}_{24}-\frac{i}{2}\Omega_{c}(\mathcal {G}_{22}-\mathcal{G}_{34}-
     \mathcal {G}_{44})-\frac{i}{2}\Omega_{p}\mathcal{G}_{21},\\ \nonumber
\dot{\mathcal{G}}_{34}&=&i\omega\mathcal{G}_{34}-\Gamma_{34}(\mathcal{G}_{33}+
     \mathcal{G}_{44})-(\Gamma_3+\Gamma_4)\mathcal{G}_{34}-\frac{i}{2}\Omega_{p}(\mathcal{G}_{31}-
     \mathcal {G}_{14})-\frac{i}{2}\Omega_{c}(\mathcal {G}_{32}-\mathcal{G}_{24}),\\ \nonumber
\end{eqnarray}
\end{widetext}
and the elements not included in Eq.~(\ref{eq:2lambda}) can be
obtained by using the complex conjugate relation:
$\mathcal{G}_{ij}=\mathcal{G}^{*}_{ji}$.

The EIT process in the detuning space $\Delta_p$-$\Delta_c$ is shown
in Fig.~\ref{fig:EIT1}. Panels (a) and (b) are the results of the
SGC factor $\beta=0$, and panels (c) and (d) are the results of
$\beta=1$. Panels (a) and (c) are absorption line shapes, and panels
(b) and (d) are Mandel's $Q_p$ parameters.

When $\beta=0$, there is no effect of SGC. The absorption line
shapes $I_p$ of the probe field is shown in Fig.~\ref{fig:EIT1} (a).
It can be seen the transparency signal is on the diagonal
$\Delta_p=\Delta_c$, corresponding to the two-photon absorption
process. For fixed detuning $\Delta_c$, there are usually three
peaks of $I_p$ as the function of $\Delta_p$ since two dark states
are induced in this system. Particularly, when $\Delta_c=0.5
\omega$, the central peak vanishes as the EIT occurs. Part of the
similar processes are studied experimentally in the
$^{85}\mathrm{Rb}$ atoms system \cite{gao} for particular coupling
detunings $\Delta_c$.

For the case of the SGC factor $\beta=1$, the GDCs $\gamma_{314}$
and $\gamma_{324}$ cause new EIT process. From Fig.~\ref{fig:EIT1}
(c), we can see the two new EIT channels along the $\Delta_p$-axis
for $\Delta_c=0.5 \omega$, and along the $\Delta_c$-axis for
$\Delta_p=0.5 \omega$ in absorption line shapes. Physically, the
GDCs $\gamma_{314}$ ($\gamma_{324}$) can cause destructive
interference between the transitions $| 3 \rangle \rightarrow | 1(2) \rangle$ and
$| 4 \rangle \rightarrow | 1(2) \rangle$ if the states
$| 3 \rangle $ and $ | 4 \rangle$ are equally excited. The dark
state is induced as the destructive interference between states $| 3
\rangle$ and $| 4 \rangle$. This can also be viewed as the
phenomenon of dark resonance \cite{car1,peng,dwang}.

These two new EIT processes are not the same with the usual EIT (the
diagonal channel) because they are assisted by SGC. We name them as
the ``vacuum-assisted transparency'' (VAT). Experimentally, for
example, we can introduce another field $\Omega_0$ between the
excited states $| 3 \rangle$ and $| 4 \rangle$ to simulate the
effects of SGC, the quantum system of the $^{87}\mathrm{Rb}$ atom
then seems to be ``locked'' by this field and the coupling field
with $\Delta_c=0.5\omega$, and the probe field can not disturb the
$^{87}\mathrm{Rb}$ atom and form the totally transparent window.
Also, we can control the transparency window by whether turning the
field $\Omega_0$ on or by changing its strength.

To understand the dynamics in this process, as an example, we
analyze the VAT channel along the $\Delta_p$-axis for $\Delta_c=0.5
\omega$. When $\Delta_c=0.5\omega$, for any $\Delta_p$ there is no
absorption of the probe field. The reason is that the population
trapping happens. When the atom is excited (except when the detuning
$\Delta_p=0.5 \omega$), the population $\rho_{11}$ transfers to
other three states, and they can not jump back to the ground state
again due to influences of SGC. The trapping is controlled by the
coupling field and GDCs. Specially, the populations $\rho_{33}$ and
$\rho_{44}$ ($\rho_{33}=\rho_{44}$)  are in proportion to the
strengths of coupling field $\Omega_c$, while the population
$\rho_{22}$ is in inverse proportion to $\Omega_c$. The process for
$\Delta_p=0.5 \omega$ is different since the usual EIT channel is
induced. The population is almost trapped in the initial ground
state ($| 1\rangle$ in our calculation). The difference between the
processes of EIT and VAT means that we can control the population
transfer \cite{Shore} by turning the detuning of the coupling field
$\Delta_c$. This process of population trapping is similar with the
studies in Refs.~\cite{zhusy2,berman,agarwal2}. In the four-level
system in Refs.~\cite{zhusy2,berman,agarwal2}, two close-lying
excited states couple to another auxiliary state by a driving field,
and decay to the single ground state. In our system, the difference
is that the coupling field and probe field can perform the EIT
channel, we can combine the effects of EIT and SGC together, and
switch between EIT ($\beta = 0$) and VAT ($\beta \ne 0$). In
addition, if the coupling field is turned off, there will be no
transparency even though SGC exists. That is to say, the coupling
field is necessary to induce the VAT channel.

The characters of $Q_p$ parameter are plotted in panels (b) and (d)
of Fig.~\ref{fig:EIT1}. As we know, when $Q>0$, the behavior of
photon statistics is bunching, and when $Q<0$, the behavior of
photon statistics is anti-bunching. In Fig.~\ref{fig:EIT1} (b), the
$Q_p$ parameter shows the standard transition between bunching and
anti-bunching when absorption occurs \cite{hey,zhengj}. When
$\beta=1$, however, the photon statistics shows bunching effect. The
fluctuation of photon emission is enhanced, especially for the two
bright branches on the $Q_p$ spectra as shown in Fig.~\ref{fig:EIT1}
(d). This means that the natures of photon emission statistics can
be greatly affected by SGC.

\subsection{$N$-TYPE SYSTEM}
\label{sec:N-type SYSTEM}

In previous section we investigate the vacuum-assisted transparency.
If we apply the switching field, the double $\Lambda$-type system
becomes the $N$-type system. The schematic diagram of this $N$-type
four-level system is shown in panel (b) of Fig.~\ref{fig.schematic}.
We study the GKE in $N$-type four-level system in this subsection,
and we compare with the experimental results of Ref.~\cite{minyan}.
Some interesting properties, such as the photon switching, frequency
conversion etc are investigated by employing this system
\cite{schmidt,myin,minyan,kang,braje,jiang,bajcsy,shujing,sinclair1,sinclair2,gong,imamoglu,gao0}.
It is shown that the process of photon emission is different with
that in the double $\Lambda$ system: the nonlinear Kerr absorption
occurs while the stability of the dark state within the EIT is
destroyed. With the effects of SGC, however, the destroyed dark
state can be repaired under some conditions.

For this system, based on the $^{87}\mathrm{Rb}$ atom, we set the
states as $|5^2S_{1/2}, F=1 \rangle = | 1 \rangle$, $|5^2S_{1/2},
F=2 \rangle = | 2 \rangle$, $|5^2P_{1/2}, F'=1 \rangle = | 3
\rangle$, and $|5^2P_{3/2}, F'=3 \rangle = | 4 \rangle$. The weak
probe field is set on transition $| 3 \rangle \rightarrow | 1
\rangle$, the strong coupling field is set on transition $| 3
\rangle \rightarrow | 2 \rangle$, the third switching field is set
on transition $| 4 \rangle \rightarrow | 2 \rangle$. The Rabi
frequencies are noted as $\Omega_{13}=\Omega_p$,
$\Omega_{23}=\Omega_c$, $\Omega_{24}=\Omega_s$. The detuning
frequencies are defined as $\Delta_p=\omega_p-\omega_{31}$,
$\Delta_c=\omega_c-\omega_{32}$, $\Delta_s=\omega_s-\omega_{42}$.
After these preparations, the equations of generating function of
Eq.~(\ref{eq:gf}) for the $N$-type system are written as follows
\begin{widetext}
\begin{eqnarray}
\label{eq:n4}
\dot{\mathcal{G}}_{11}&=&2s(\gamma_{31}\mathcal{G}_{33}+\gamma_{314}\mathcal{G}_{34}+
      \gamma_{314}\mathcal{G}_{43}+\gamma_{41}\mathcal{G}_{44})
      -\frac{i}{2}\Omega_{p}(\mathcal{G}_{13}-\mathcal{G}_{31}), \\ \nonumber
\dot{\mathcal{G}}_{22}&=&2(\gamma_{32}\mathcal{G}_{33}+\gamma_{324}\mathcal{G}_{34}+
      \gamma_{324}\mathcal{G}_{43}+\gamma_{42}\mathcal{G}_{44})
      -\frac{i}{2}\Omega_{c}(\mathcal {G}_{23}-\mathcal{G}_{32})-
      \frac{i}{2}\Omega_{s}(\mathcal{G}_{24}-\mathcal {G}_{42}),\\ \nonumber
\dot{\mathcal{G}}_{33}&=&-2\Gamma_3\mathcal{G}_{33}-\Gamma_{34}(\mathcal{G}_{34}+\mathcal{G}_{43})
      -\frac{i}{2}\Omega_{p}(\mathcal{G}_{31}-\mathcal {G}_{13})-
      \frac{i}{2}\Omega_{c}(\mathcal {G}_{32}-\mathcal{G}_{23}),\\ \nonumber
\dot{\mathcal{G}}_{44}&=&-2\Gamma_4\mathcal{G}_{44}-\Gamma_{34}(\mathcal{G}_{34}+\mathcal{G}_{43})
      -\frac{i}{2}\Omega_{s}(\mathcal{G}_{42}-\mathcal{G}_{24}),\\ \nonumber
\dot{\mathcal{G}}_{12}&=&-i(\Delta_p-\Delta_c)\mathcal{G}_{12}-
      \frac{i}{2}(\Omega_{c}\mathcal{G}_{13}+\Omega_{s}\mathcal{G}_{14}-
      \Omega_{p}\mathcal{G}_{32}), \\ \nonumber
\dot{\mathcal{G}}_{13}&=&-i\Delta_p\mathcal{G}_{13}-
      \Gamma_3\mathcal{G}_{13}-\Gamma_{34}\mathcal{G}_{14}
      -\frac{i}{2}\Omega_{p}(\mathcal{G}_{11}-\mathcal{G}_{33})-
      \frac{i}{2}\Omega_{c}\mathcal {G}_{12},\\ \nonumber
\dot{\mathcal{G}}_{14}&=&-i(\Delta_p-\Delta_c+\Delta_s)\mathcal{G}_{14}-
      \Gamma_{34}\mathcal{G}_{13}-\Gamma_4\mathcal{G}_{14}
      -\frac{i}{2}(\Omega_{s}\mathcal{G}_{12}-\Omega_{p}\mathcal{G}_{34}),\\ \nonumber
\dot{\mathcal{G}}_{23}&=&-i\Delta_c\mathcal{G}_{23}-\Gamma_3\mathcal{G}_{23}-
      \Gamma_{34}\mathcal{G}_{24}-\frac{i}{2}\Omega_{c}(\mathcal {G}_{22}-\mathcal{G}_{33})-
      \frac{i}{2}(\Omega_{p}\mathcal{G}_{21}-\Omega_{s}\mathcal {G}_{43}),\\ \nonumber
\dot{\mathcal{G}}_{24}&=&-i\Delta_s\mathcal{G}_{24}-\Gamma_{34}\mathcal{G}_{23}-
      \Gamma_4\mathcal{G}_{24}-\frac{i}{2}\Omega_{s}(\mathcal {G}_{22}-
      \mathcal {G}_{44})+\frac{i}{2}\Omega_{c}\mathcal{G}_{34},\\ \nonumber
\dot{\mathcal{G}}_{34}&=&-i(\Delta_s-\Delta_c)\mathcal{G}_{34}-
     \Gamma_{34}(\mathcal{G}_{33}+\mathcal{G}_{44})-(\Gamma_3+\Gamma_4)\mathcal{G}_{34}
     -\frac{i}{2}(\Omega_{s}\mathcal{G}_{32}-\Omega_{p}\mathcal {G}_{14}-
     \Omega_{c}\mathcal{G}_{24}),\\ \nonumber
\end{eqnarray}
\end{widetext}
and the other elements not included in Eq.~(\ref{eq:n4}) satisfy the
complex conjugate relation: $\mathcal{G}_{ij}=\mathcal{G}^{*}_{ji}$.

\begin{figure}
\includegraphics[scale=0.25]{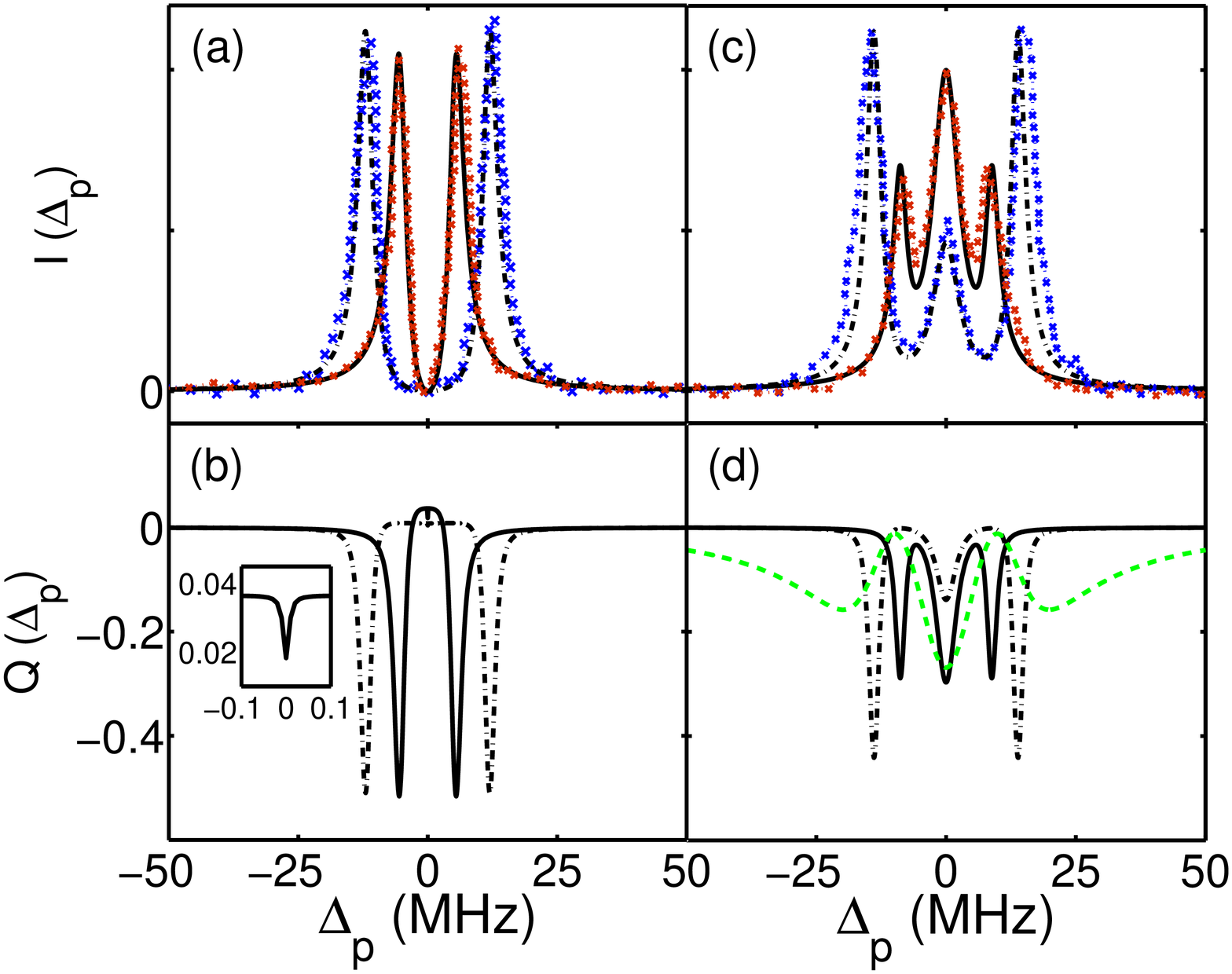}
\caption{(Color online) The absorption line shapes $I_p$ (panels (a)
and (c)) and Mandel's $Q_p$ parameter (panels (b) and (d)) of the
$N$-type system. Panels (a) and (b) is for the switching field
turn-off, panels (c) and (d) is for the switching field turn-on,
$\Omega_s=14$MHz. The blue and red crosses are the experimental
results in Fig. $3$ of Ref.~\cite{minyan}. The inset shows the fine
structure of the hole on the $Q_p$ spectra for the solid line around
$\Delta_p=0$. The parameters are
$\gamma_{31}=\gamma_{32}=1.4375$MHz,
$\gamma_{41}=\gamma_{42}=1.5167$MHz. $\Delta_c$=$\Delta_s=0$, and
$\beta=0$. $\Omega_p=1.5$MHz, $\Omega_c=11$MHz (solid lines),
$\Omega_c=24$MHz (dashed-dot lines). The initial condition is
$|\psi\rangle$=$| 1 \rangle$. The dashed green line of Mandel's
$Q_p$ parameter is under the conditions of $\Omega_p=15$MHz,
$\Omega_c=24$MHz, $\Omega_s=14$MHz. } \label{fig:KE1}
\end{figure}

\begin{figure}
\includegraphics[scale=0.2]{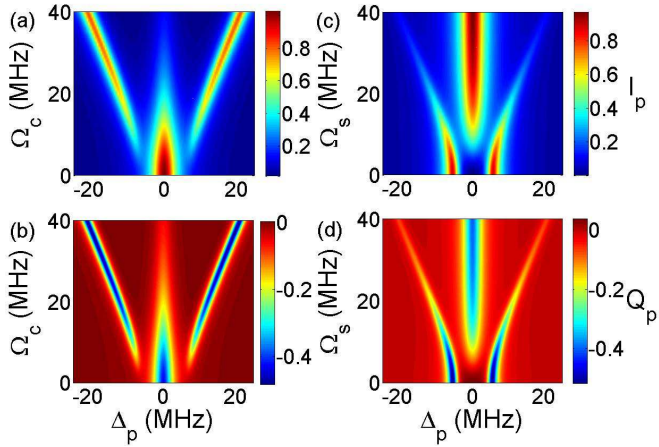}
\caption{(Color online) The absorption line shapes $I_p$ (panels (a)
and (c)) and Mandel's $Q_p$ parameter (panels (b) and (d)) of the
$N$-type system as the functions of Rabi frequencies $\Omega_c$
(panels (a) and (b)), $\Omega_s$ (panels (c) and (d)) and detuning
frequency $\Delta_p$. The parameters are $\Omega_p=1.5$MHz,
$\Omega_s=14$MHz (panels (a) and (b)); $\Omega_p=1.5$MHz,
$\Omega_c=11$MHz (panels (c) and (d)). The other parameters are the
same with Fig.~\ref{fig:KE1}. } \label{fig:KE2}
\end{figure}

\begin{figure}%[b]
\includegraphics[scale=0.25]{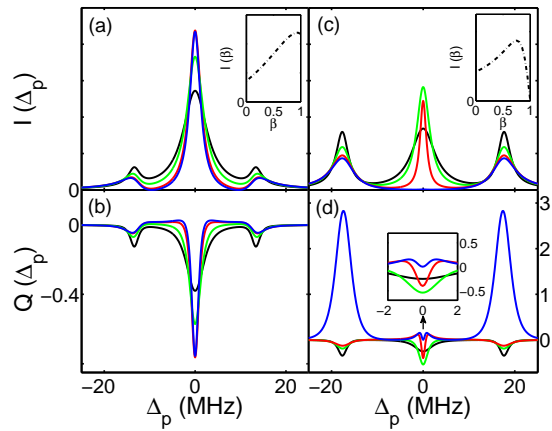}
\caption{(Color online) The absorption line shapes $I_p$ (panels (a)
and (c)) and Mandel's $Q_p$ parameter (panels (b) and (d)) of the
$N$-type system. $\Omega_p=1.5$MHz, $\Omega_s=25$MHz,
$\Omega_c=10$MHz (panels (a) and (b)), $25$MHz (panels (c) and (d)).
The SGC factor $\beta=0$ (black line), $\beta=0.5$ (green line),
$\beta=0.9$ (red line), $\beta=1$ (blue line). The other parameters
are the same with Fig.~\ref{fig:KE1}. The insets in panels (a) and
(c) show the height of the Kerr nonlinear absorption peak of line
shapes as the function of $\beta$, the inset in panel (d) shows
Mandel's $Q_p$ parameter near resonance $\Delta_p=0$.}
\label{fig:KE3}
\end{figure}

The results of the GKE for weak probe field are shown in
Figs.~\ref{fig:KE1}, \ref{fig:KE2}, and \ref{fig:KE3}.

In Fig.~\ref{fig:KE1}, we show the EIT and the GKE in this system.
Panels (a) and (c) are absorption line shapes $I_p$,  panels (b) and
(d) are $Q_p$ parameter. The blue and red crosses are the
experimental results of Ref.~\cite{minyan}, the solid lines and
dashed-dot lines are our theoretical results. The experiment of
Ref.~\cite{minyan} is carried out in a vapor-cell magneto-optic trap
(MOT). The trapped Rb atom clouds is about $1$mm in diameter and
contains about $2\times 10^7$ Rb atoms. The Rb atom vapor pressure
is about $10^{-8}$Torr. In the experiment of Ref.~\cite{minyan},
there is no effect of SGC, that is, the SGC factor $\beta=0$. As
shown in the figure, our theoretical results and experimental
results are in agreement well \cite{note}. Fig.~\ref{fig:KE1} (a)
and (b) show the standard EIT signal when the switching field is
turned off. As the Rabi frequency $\Omega_c$ increases from $11$MHz
to $24$MHz, the absorption line shape $I_p$ and $Q_p$ spectra show
the power broaden, which reflects the influence of the strengths of
the coupling filed. The behavior of the photon emission at resonance
shows anti-bunching. In contrast, when the EIT occurs at
$\Delta_p=0$, the behavior of photon emission shows bunching. And,
there is the sharp ``hole'' in $Q_p$ spectra which is shown in the
inset of Fig.~\ref{fig:KE1} (b). The reason for this ``hole'' is
that the photon emission stops once the dark state is formed
\cite{peng,dwang}.

When the switching field $\Omega_s$ is turn-on, there is absorption
at $\Delta_p=0$ as shown in Fig.~\ref{fig:KE1} (c). The two
absorption sidebands are the usual Aulter-Townes (AT) doublets, and
the central peak is the Kerr nonlinear absorption. There is no
``hole'' on the $Q_p$ spectra since the EIT is destroyed in this
process. We show $Q_p$ parameter for the strengths of probe field
$\Omega_p=15$MHz by using dashed green line in Fig.~\ref{fig:KE1}
(d). This result is not the same with those under the usual
condition where anti-bunching behavior can not be captured as the
strengths of the probe field increases \cite{peng,dwang}. And the
anti-bunching behavior of photon emission holds on as the strengths
of the probe field $\Omega_p$ increases. In other words, the
strengths of the probe field can not change the anti-bunching
behavior. This stable anti-bunching behavior of photon emission
demonstrates the ability of photon switching, and that the $N$-type
system based on the GKE can serve as the few photon emitter
\cite{imamoglu2}.

Furthermore, we show the absorption line shapes $I_p$ (panels (a)
and (c)) and $Q_p$ parameter (panels (b) and (d)) as the function of
Rabi frequencies $\Omega_c$ (panels (a) and (b)), and $\Omega_s$
(panels (c) and (d)) in Fig.~\ref{fig:KE2}. It shows that the
strengths of the driven fields play an important role in the GKE.
They can control the nonlinearity and the AT splitting.

As shown in Fig.~\ref{fig:KE2} (a) there is mainly one central peak for the absorption
line shapes at $\Delta_p=0$ for small $\Omega_c$,
corresponding to the one-photon process $| 3 \rangle \rightarrow | 1 \rangle$.
From the view of dressed state theory \cite{minyan,gao0}, the
coupling field with ``pure'' state $| 3 \rangle$ induce a pair of dressed
states $| - \rangle$ and $| + \rangle$ (instead of state $| 3 \rangle$).
The transitions from states $| 1 \rangle$ to $| - \rangle$ and $| + \rangle$ are
probed by the probe field $\Omega_p$, the transitions
from states $| - \rangle$ and $| + \rangle$ to $| 4 \rangle$ are
pumped by the switching field $\Omega_s$. The separation between states $| - \rangle$ and $| + \rangle$
are controlled via the coupling field.
When $\Delta_c=0$, the two dressed states are equally
driven, the height of AT doublets are the same. As the Rabi
frequency $\Omega_c$ increases, the one-photon process is weakened,
that is, the interference between the transitions $| - \rangle \rightarrow | 1 \rangle$
and $| + \rangle \rightarrow | 1 \rangle$
is destructive. In contrast, the two-photon processes
$| 4 \rangle \rightarrow | - \rangle \rightarrow | 1 \rangle$ and
$| 4 \rangle \rightarrow | + \rangle \rightarrow | 1 \rangle$ are being enhanced,
and the height of AT doublets at sidebands gets bigger.

For small $\Omega_s$ there is only AT doublets of absorption line
shapes $I_p$, this is not the same with the above case. It is shown
in Fig. \ref{fig:KE2} (c). Under this condition, the Kerr
nonlinearity is trivial, which can also be realized for far off
resonance of the switching field. As $\Omega_s$ increases, however,
the central peak at $\Delta_p=0$ increases nonlinearly until its
steady value. Physically, this nonlinear increase is determined by
the strengths of the coupling field $\Omega_c$, and the steady
process of $I_p$ is related to the probe field. Another influence of
switching field is, the splitting of the AT doublets gets bigger
when $\Omega_s$ increases. The reason is that the splitting between
the dressed states can also be affected by the switching field
$\Omega_s$, this splitting is $\propto
\sqrt{\Omega_c^2+\Omega_s^2}$. Our numerical results verify this
behavior.

In Fig.~\ref{fig:KE2} (b) and (d) we show the spectra of Mandel's
$Q_p$ parameters. For the central Kerr absorption at $\Delta_p=0$,
one can see magnitude of $Q_p$ parameters decreases with $\Omega_c$
and increases with $\Omega_s$. The reason is that the coupling field
can inhibit the Kerr absorption process, while the switching field
can enhance it. However, $Q_p$ parameters stay minus at resonance,
which means the strengths of the coupling field $\Omega_c$ or the
switching field $\Omega_s$ can not change the natures of the photon
statistics: the anti-bunching or the bunching.

In the above consideration we do not include the effect of SGC. In
fact, it is also possible to introduce the effect of SGC into the
nonlinear Kerr absorption. The results for different SGC factor
$\beta$ are shown in Fig.~\ref{fig:KE3}. The top row is absorption
line shapes $I_p$, and the bottom row is $Q_p$ parameter.

To better show GKE, we chose the parameters to make the one-photon
process obvious. In Fig.~\ref{fig:KE3} (a) and (b), we show the
results of $\Omega_s=25$MHz, $\Omega_c=10$MHz. We can see from the
figure, the Kerr absorption is more obvious than that of AT
doublets. As the SGC factor $\beta$ increases (from $0$ to $1$), the
AT doublets are being restrained. The central Kerr peak, however, is
``coherently narrowed and enhanced'' (see Fig.~\ref{fig:KE3} (a)).
The behavior of the Kerr absorption shows nonlinear change with the
increasing of $\beta$, as shown in the inset of Fig.~\ref{fig:KE3}
(a). The values of $Q_p$ parameters (panel (b)) become negative big
at $\Delta_p=0$ as $\beta$ increases, which means that the
anti-bunching effect becomes more obvious. Also, $Q_p$ spectra
becomes narrow around $\Delta_p=0$.

We further consider another special case: $\Omega_c=\Omega_s=25$MHz.
The results are shown in Fig.~\ref{fig:KE3} (c) and (d). We see that
the separation of the AT doublets increases to $35.36$MHz. The Kerr
absorption peak as the function of $\beta$ is shown in the inset of
Fig.~\ref{fig:KE3} (c). Its behavior in this case is different with
that of the above case. What is particular in this case is the
central Kerr absorption goes to zero when $\beta=1$. That is, the
nonlinear Kerr absorption is eliminated due to SGC, it is shown as
the blue line in Fig.~\ref{fig:KE3} (c). This can also be viewed as
the retrieval of dark state in the system. The reason is that there
exists the quantum interference between the one-photon and
two-photon processes for $\beta=1$. For $\beta=0$, however, there is
no this quantum interference. In addition, the behavior of photon
emission is affected by SGC: $Q_p$ parameter changes from negative
to positive values when $\beta$ gets big, which is different with
the above case. The inset in panel (d) shows this change at the
detuning $\Delta_p=0$.\\

\section{BRIEF CONCLUSION}
\label{sec:CL}

In this work we investigate the effects of SGC on the
electromagnetically induced transparency (EIT) and the giant Kerr
effect (GKE) of four-level quantum system of the $^{87}\mathrm{Rb}$ atom
driven by external cw laser fields.

In our study, we show the structure of absorption line shapes in the
detuning space for the double $\Lambda$-type system. The usual EIT
means that we can control the quantum system (such as the
$^{87}\mathrm{Rb}$ atom) by employing the strong coupling field, and
there is no absorption when the weak probe field is introduced under
the condition of two-photon resonance. With the effect of SGC, there
exist new EIT channels, we note them as the ``vacuum-assisted
transparency" (VAT). It means that we can use external fields to
lock the system, so that there is no absorption when the probe field
is introduced even beyond the two-photon resonance condition. The
transparency window crosses the whole detuning range of the probe
field. Also, this could be more convenient to operate in experiment
than the usual EIT setup.

The coupling field $\Omega_c$ and the switching field $\Omega_s$
play the different role in GKE for the $N$-type system: the former
one can inhibit the nonlinear absorption, while the later one can
enhance it. If there exists SGC in the system, the nonlinear Kerr
absorption can be either narrowed or eliminated under different
driven conditions. When the nonlinear absorption is completely
eliminated, the dark state can be repaired by SGC.

For the photon statistics, several kinds of behaviors of Mandel's
$Q$ parameter, such as the ``hole'' of the dark state, the steady
anti-bunching of the photon switching, the transition between
bunching and anti-bunching near resonance, and the big fluctuation
due to SGC are also studied.

Based on this work, the dynamics of other type four level systems,
such as the $Y$-type, the inverse $Y$-type systems can be
investigated. Also, this study will be of interest for further
investigation in few photon processes, spontaneous emission
cancellation, etc.

\begin{acknowledgments}
We thank the referee for a helpful comments.
This work was supported by the National Science Foundation of China
(Grant No. 91021009, 10874102), and National Basic Research Program
of China (973 Program, Grant No. 2009CB929404).
\end{acknowledgments}


\begin{thebibliography}{99}
\bibitem{note} The tiny departure is mainly due to the atom collision and finite laser
            linewidths in practice as pointed out in Ref.~\cite{minyan}.
\end{thebibliography}
\end{document}